\documentstyle[aps,prl,epsf,epsfig,floats,twocolumn]{revtex}
\begin{document}
\draft
\twocolumn[\hsize\textwidth\columnwidth\hsize\csname @twocolumnfalse\endcsname

\title{Magnetotransport in the doped Mott insulator}
\author{Ekkehard Lange and Gabriel Kotliar}
\address{Serin Physics Laboratory,
Rutgers University,
136 Frelinghuysen Road,
Piscataway, New Jersey 08854}
%\date{\today}
\maketitle
\begin{abstract}
We investigate  the Hall effect and the magnetoresistance of strongly
correlated electron systems using the dynamical mean-field theory. We
treat the low- and high-temperature limits analytically and  explore
some aspects of the intermediate-temperature regime numerically. We
observe that a bipartite-lattice condition is responsible for the
high-temperature result $\sigma_{xy}\sim1/T^2$ obtained by various
authors, whereas the general behavior is $\sigma_{xy}\sim1/T$, as for
the longitudinal conductivity. We find that Kohler's rule is neither
obeyed at high nor at intermediate temperatures. 
\end{abstract}
%%
%%71.27.+a		Strongly correlated electron systems
%%72.15.Gd		magnetotransport effects
%%75.20.Hr		Local moment, Kondo effect, valence fluct., HF.
%%74.20.Mn		Nonconventional mechanisms
%%
]

\section{Introduction}
\label{intro}

The Hubbard model \cite{Hubbard:1963,Gutzwiller:1963,Kanamori:1963} of
strongly correlated electron systems has been an enduring problem in
condensed matter theory. It is believed to capture some of the
anomalous physics of heavy-fermion systems and high-$T_c$
superconductors \cite{Fulde}. Its crucial feature is the interplay of
itineracy and a local interaction $U$ that is either comparable or
much greater than the bare bandwidth. 

In this paper, we investigate the impact of a magnetic field on the
charge transport in strongly correlated electron systems within the
Hubbard model. This issue involves
different quantities that are closely related and should therefore be
considered within the {\it same} approximation scheme. We meet this
condition by using the dynamical mean-field theory which becomes exact
in the limit of infinite dimensions. A major though nontrivial
simplification of this approach is that transport properties are
described solely by the single-particle spectrum
\cite{Mueller-Hartmann:1989a,Khurana:1990,Georges:1996,Pruschke:1995}.
While the dynamical mean-field theory still captures many properties
of real three-dimensional transition-metal oxides, it fails to describe
cuprate superconductors equally well, mainly because it does not
properly take into account magnetic correlations. Nevertheless, it is
important to fully explore this approximation scheme in order to
establish a sound starting point for future improvements.

The Hall constant of the single-band Hubbard model has already been
considered within the dynamical mean-field theory by Pruschke {\it et
al.} \cite{Pruschke:1995} and  Majumdar and Krishnamurthy
\cite{Majumdar:1995}. The former authors have computed the Hall
constant and Hall angle as a function of temperature for various
doping levels at --- in our units --- $U=2\sqrt{2}D$ ($D$ is the half
bandwidth) using the non-crossing approximation (NCA) and a quantum
Monte Carlo (QMC) technique to solve the single-impurity
problem. Majumdar and Krishnamurthy have mainly focused on the
relation between the infinite-frequency Hall constant investigated by
Shastry, Shraiman and Singh \cite{Shastry:1993} and the dc Hall
constant, using the iterated-perturbation theory (IPT) of Ref.\
\cite{Kajueter:1996b}.

In addition to the Hall effect and the ordinary resistivity, we
investigate the magnetoresistance. We tie our numerical analysis at
intermediate temperatures to analytical results valid in the low- and
high-temperature limits. This allows to disentangle the coherent and 
incoherent contributions of the single-particle spectrum to the 
magnetotransport and to gain some understanding of either. We always 
mainly focus on the parameter regime close to the density-driven Mott 
transition. This paper is organized as follows: In Sec.\ \ref{theory},
we briefly summarize the dynamical mean-field theory within the 
single-band Hubbard model and derive expressions for the quantities to
be calculated. Then, by using a  Fermi-liquid parametrization for the 
spectral function, we examine all quantities in the low-temperature 
regime (Sec.\ \ref{lowTlimit}) and study their dependences on 
temperature and on the doping level. Moreover, we discuss the impact 
of correlations close to half filling. In Sec.\ \ref{highTlimit}, we 
employ a recently developed scheme to expand transport coefficients in
powers of $1/T$ \cite{Gunnar} within the $U=\infty$ Hubbard model to 
study the opposite limit of high temperatures. We show that the
high-temperature behaviors of the Hall constant and Hall angle can be
affected by specific lattice symmetries and that the Hall angle
increases at most linearly with temperature. Then, we explore the
intermediate-temperature regime numerically (Sec.\ \ref{intTregime})
by either using the NCA or the IPT, depending on whether our focus is
more on higher or lower temperatures, respectively. Finally, in
Sec.\ \ref{conclusions}, we summarize and discuss our results. 

%%%%%%%%%%%%%%%%%%%%%%%%%%%%%%%%%%%%%%%%%%%%%%%%%%%%%%%%%%%%%%%%%%%%%%%%%%%%%
%%%%%%%%%%%%%%%%%%%%%%%%%%%%%%%%%%%%%%%%%%%%%%%%%%%%%%%%%%%%%%%%%%%%%%%%%%%%%
%%%%%%%%%%%%%%%%%%%%%%%%%%%%%%%%%%%%%%%%%%%%%%%%%%%%%%%%%%%%%%%%%%%%%%%%%%%%%
%%%%%%%%%%%%%%%%%%%%%%%%%%%%%%%%%%%%%%%%%%%%%%%%%%%%%%%%%%%%%%%%%%%%%%%%%%%%%
%%%%%%%%%%%%%%%%%%%%%%%%%%%%%%%%%%%%%%%%%%%%%%%%%%%%%%%%%%%%%%%%%%%%%%%%%%%%%
%%%%%%%%%%%%%%%%%%%%%%%%%%%%%%%%%%%%%%%%%%%%%%%%%%%%%%%%%%%%%%%%%%%%%%%%%%%%%
%%%%%%%%%%%%%%%%%%%%%%%%%%%%%%%%%%%%%%%%%%%%%%%%%%%%%%%%%%%%%%%%%%%%%%%%%%%%%
%%%%%%%%%%%%%%%%%%%%%%%%%%%%%%%%%%%%%%%%%%%%%%%%%%%%%%%%%%%%%%%%%%%%%%%%%%%%%
%%%%%%%%%%%%%%%%%%%%%%%%%%%%%%%%%%%%%%%%%%%%%%%%%%%%%%%%%%%%%%%%%%%%%%%%%%%%%
%%%%%%%%%%%%%%%%%%%%%%%%%%%%%%%%%%%%%%%%%%%%%%%%%%%%%%%%%%%%%%%%%%%%%%%%%%%%%
%%%%%%%%%%%%%%%%%%%%%%%%%%%%%%%%%%%%%%%%%%%%%%%%%%%%%%%%%%%%%%%%%%%%%%%%%%%%%
%%%%%%%%%%%%%%%%%%%%%%%%%%%%%%%%%%%%%%%%%%%%%%%%%%%%%%%%%%%%%%%%%%%%%%%%%%%%%
%%%%%%%%%%%%%%%%%%%%%%%%%%%%%%%%%%%%%%%%%%%%%%%%%%%%%%%%%%%%%%%%%%%%%%%%%%%%%
%%%%%%%%%%%%%%%%%%%%%%%%%%%%%%%%%%%%%%%%%%%%%%%%%%%%%%%%%%%%%%%%%%%%%%%%%%%%%
%%%%%%%%%%%%%%%%%%%%%%%%%%%%%%%%%%%%%%%%%%%%%%%%%%%%%%%%%%%%%%%%%%%%%%%%%%%%%
%%%%%%%%%%%%%%%%%%%%%%%%%%%%%%%%%%%%%%%%%%%%%%%%%%%%%%%%%%%%%%%%%%%%%%%%%%%%%

\section {Formalism}
\label{theory}

We consider the $N_s$-fold degenerate Hubbard model,
\begin{equation}
 H=-t\sum_{\langle ij\rangle\sigma}c_{i\sigma}^{+}c_{j\sigma}
 +{U \over 2}\sum_{i\sigma\neq\sigma'}
 n_{i\sigma}n_{i\sigma'},
\label{Hubbard}
\end{equation}
where in the first term, the sum is over nearest neighbors. The index
$\sigma$ can be thought of as a spin or an orbital index, and will run
from 1 to $N_s$. For a single band, $N_s=2$. In the limit of infinite
spatial dimensions, $d\rightarrow\infty$, the irreducible self-energy
and all vertex functions collapse onto a single site
\cite{Mueller-Hartmann:1989a,Metzner:1989b}. As a consequence, these
functions do no longer depend on momentum. This, in turn, implies that
all vertex corrections of the conductivity tensor
vanish\cite{Khurana:1990,Georges:1996,Pruschke:1995}, which therefore
can be calculated from the single-particle spectral function
\begin{equation}
 A(\omega,\epsilon_{\vec{k}})=-{1\over\pi}\Im
	G(\omega,\epsilon_{\vec{k}}).
\label{spectral}
\end{equation}
Here, $G(\omega,\epsilon_{\vec{k}})$ is the retarded Green's function
\begin{equation}
 G(\omega,\epsilon_{\vec{k}})=\frac{1}
	{\omega+\mu-\epsilon_{\vec{k}}-\Sigma(\omega)},
\label{Green}
\end{equation}
where $\mu$ is the chemical potential and $\Sigma(\omega)$ is the
momentum-independent self-energy. The Green's function (\ref{Green})
must be calculated by solving a single-impurity Anderson model
supplemented by a self-consistency condition \cite{Georges:1996}. In
terms of the spectral function (\ref{spectral}), the ordinary
conductivity, the Hall conductivity, and the magnetoconductance read
\cite{Voruganti:1992}: 
\begin{eqnarray}
 \sigma_{xx}&=&N_s\pi e^2\int d\epsilon\,\phi_{xx}(\epsilon)
 	\int d\omega\,[-\frac{\partial f(\omega)}{\partial\omega}]
	A(\omega,\epsilon)^2,\label{condxx}\\
 \sigma_{xy}&=&\frac{N_s 2\pi^2e^3H}{3}\int
	       d\epsilon\,\phi_{xy}(\epsilon)
	\int d\omega\,[-\frac{\partial f(\omega)}{\partial\omega}]
	A(\omega,\epsilon)^3,\label{condxy}\\
 \Delta\sigma_{xx}&=&\frac{N_s2\pi^3e^4H^2}{5}\int d\epsilon\,
               \phi_{M}(\epsilon)
	\int d\omega\,[-\frac{\partial f(\omega)}{\partial\omega}]
	A(\omega,\epsilon)^4\label{condM}.
\end{eqnarray}
Here, $H$ is the magnetic field and $e$ denotes the charge of an
electron, hence, $e<0$. Furthermore,
$f(\omega)=1/[\mbox{exp}(\beta\omega)+1]$ is the Fermi function, where 
$\beta=1/T$ is the inverse temperature. Since the spectral function
depends on $\vec{k}$ only via the band dispersion
$\epsilon_{\vec{k}}$, we could write all sums over the Brillouin zone
as integrals over the following transport functions
\begin{eqnarray}
 \phi_{xx}(\epsilon)&=&{1\over N}\sum_{\vec{k}}(\epsilon_{\vec{k}}^x)^2
	\delta(\epsilon-\epsilon_{\vec{k}}),\label{phixx}\\
 \phi_{xy}(\epsilon)&=&{1\over N}\sum_{\vec{k}}\mbox{det}(\vec{k})
	\delta(\epsilon-\epsilon_{\vec{k}}),\label{phixy}\\
 \phi_{M}(\epsilon)&=&{1\over N}\sum_{\vec{k}}M(\vec{k})
	\delta(\epsilon-\epsilon_{\vec{k}}).\label{phiM} 
\end{eqnarray}
Here, $N$ denotes the total number of lattice sites. Upper indices
indicate differentiations with respect to a component of the Bloch
vector such as in, say,
\begin{equation}
 \epsilon_{\vec{k}}^x={\partial\epsilon_{\vec{k}}\over\partial k_x}.
\end{equation}
Finally, the integrands of Eqs.\ (\ref{phixy}) and (\ref{phiM})
contain the functions
\begin{eqnarray}
 \mbox{det}(\vec{k})&=&
   \left|\begin{array}{cc}
	\epsilon_{\vec{k}}^x\epsilon_{\vec{k}}^x &
	\epsilon_{\vec{k}}^{xy}\\[1.4 ex]
	\epsilon_{\vec{k}}^y\epsilon_{\vec{k}}^x &
	\epsilon_{\vec{k}}^{yy}
   \end{array}\right|\label{detk},\\
 M(\vec{k})&=&\epsilon_{\vec{k}}^{xxx}\epsilon_{\vec{k}}^{x}
 	      (\epsilon_{\vec{k}}^{y})^2
	-2\epsilon_{\vec{k}}^{xxy}(\epsilon_{\vec{k}}^{x})^2
	      \epsilon_{\vec{k}}^{y}
	+\epsilon_{\vec{k}}^{xyy}(\epsilon_{\vec{k}}^{x})^3
\nonumber\\
 	&-&(\epsilon_{\vec{k}}^{x})^2
	      [\epsilon_{\vec{k}}^{xx}\epsilon_{\vec{k}}^{yy}
	      -(\epsilon_{\vec{k}}^{xy})^2]\label{Mk}.
\end{eqnarray}
On a hypercubic lattice in $d$ dimensions,
\begin{equation}
 \label{dispersion}
 \epsilon_{\vec{k}}=-{2t\over\sqrt{2d}}\sum_{i=1}^{d}\cos(k_{i}a).
\end{equation}
Henceforth, we set the lattice spacing $a$ and the half bandwidth
$D\equiv2t$ equal to 1. For the band of Eq.\ (\ref{dispersion}), 
the noninteracting density of states, $D(\epsilon)$, and the transport
functions (\ref{phixx})-(\ref{phiM}) can easily be calculated to be
\begin{eqnarray}
 D(\epsilon)&=&\sqrt{2/\pi}\;e^{-2\epsilon^2}\label{Gauss},\\
 \phi_{xx}(\epsilon)&=&{1\over 4d}\;D(\epsilon),\label{phixx2}\\  
 \phi_{xy}(\epsilon)&=&-{1\over 4d^2}\;\epsilon D(\epsilon),\label{phixy2}\\
 \phi_{M}(\epsilon)&=&-{1\over 16d^2}\;D(\epsilon).\label{phiM2} 
\end{eqnarray} 
Obviously, $\sigma_{xx}$, $\sigma_{xy}$, and $\Delta\sigma_{xx}$ are
of the orders $1/d$, $1/d^2$, and $1/d^2$, respectively. Consequently,
the Hall constant and the magnetoresistance are of zeroth order in
$d$ and are given by
\begin{eqnarray}
 R_H&=&{\sigma_{xy}\over\sigma_{xx}^2H}\label{resiH},\\
 \Delta\rho&=&-{\Delta\sigma_{xx}\over\sigma_{xx}^2}\label{resiM},
\end{eqnarray} 
as $d\rightarrow\infty$, respectively. Eqs.\ (\ref{condM}),
(\ref{phiM2}), and (\ref{resiM}) imply that the magnetoresistance is
strictly nonnegative in infinite dimensions. This means that the
coupling of the magnetic field to the orbital motion of the electrons
enhances the resistivity. At any finite dimension, there is an
additional contribution from the Hall conductivity to the
magnetoresistance which is of opposite sign,
$\Delta\rho=-\rho^2\left(\Delta\sigma_{xx}+\rho\sigma_{xy}^2\right)$,
where $\rho=1/\sigma_{xx}$. In this work, rather than using the
Gaussian density of states (\ref{Gauss}), we choose the semicircular
one, 
\begin{equation}
 D(\epsilon)={2\over\pi}\;\Theta(1-|\epsilon|)\sqrt{1-\epsilon^{2}},
\label{DOS}
\end{equation}
since then, the $\epsilon$ integrals in Eqs.\ (\ref{condxx})-(\ref{condM})
can be done analytically. Here, $\Theta(x)$ is the Heaviside function,
which is either 1 or 0 depending on whether $x$ is greater or smaller
than 0, respectively. 

To calculate the spectral function (\ref{spectral}), we employ two
methods: At low temperatures, we use the IPT modified for finite
doping levels as described in Ref.\ \cite{Kajueter:1996b}. This method
becomes exact in various limits at zero temperature and has been
extended to finite temperatures in Ref.\ \cite{Kajueter:thesis}. 
At high temperatures, we use the NCA for the infinite-$U$ case
\cite{Coleman:1984}. This approach has been shown to give results
which are in good agreement with both QMC \cite{Pruschke:1993b} and
numerical renormalization group calculations \cite{Costi:1996} at high
temperatures.

Before proceeding, we discuss the relevant energy scales. We are
primarily interested in the physics close to half filling,
$\delta\rightarrow0$, where $n=1-\delta$ denotes the average occupancy
per lattice site. To understand the physics in this regime, we
employ the relation of the dynamical mean-field theory to the
single-impurity Anderson model. For $\delta\ll1$, we are in the
local-moment regime and an Abrikosov-Suhl resonance shows up in the
local spectral function
\begin{equation}
\label{localspectral}
 A(\omega)=\int_{-\infty}^{\infty}d\epsilon\,D(\epsilon)A(\omega,\epsilon)
\end{equation}
for low enough temperatures. Its width defines an energy scale $T^*$
\cite{Jarrell:1993,Kajueter:1996a}, which is not to be confused with
the quasiparticle damping (\ref{damping}) that we will introduce
further down. The emergence of a resonance at the Fermi 
level indicates a Kondo-like screening of the local moment. In fact,
the local spin susceptibility crosses over from local- to
screened-moment behavior at a ``Kondo temperature'' $T_{\mbox{\small
coh}}$ \cite{Jarrell:1993,Kajueter:1996a}. $T_{\mbox{\small coh}}$
defines the coherence temperature below which Fermi-liquid theory is
applicable \cite{Kajueter:1996a}. Generally, $T_{\mbox{\small
coh}}\ll T^*$. The emergence of two low-energy scales is known from
Kondo-lattice systems \cite{Hewson}. Finally, our high-energy scale is
$D$ since we are only interested in temperatures much smaller than the
Mott-Hubbard gap, $T\ll U$.

For temperatures most pertinent to experiments, the magnetotransport
is governed by both the coherence peak and the incoherent backround of 
the spectrum. It is of fundamental interest to separate both 
contributions. In the Fermi-liquid regime, the transport is entirely 
determined by the coherence peak, and our analysis of Sec.\
\ref{lowTlimit}\ addresses its contribution only. Our high-temperature
analysis of Sec.\ \ref{highTlimit}, on the other hand, captures the
contribution of the completely incoherent lower Hubbard band to the
magnetotransport.

%%%%%%%%%%%%%%%%%%%%%%%%%%%%%%%%%%%%%%%%%%%%%%%%%%%%%%%%%%%%%%%%%%%%%%%%%%%%
%%%%%%%%%%%%%%%%%%%%%%%%%%%%%%%%%%%%%%%%%%%%%%%%%%%%%%%%%%%%%%%%%%%%%%%%%%%%
%%%%%%%%%%%%%%%%%%%%%%%%%%%%%%%%%%%%%%%%%%%%%%%%%%%%%%%%%%%%%%%%%%%%%%%%%%%%
%%%%%%%%%%%%%%%%%%%%%%%%%%%%%%%%%%%%%%%%%%%%%%%%%%%%%%%%%%%%%%%%%%%%%%%%%%%%
%%%%%%%%%%%%%%%%%%%%%%%%%%%%%%%%%%%%%%%%%%%%%%%%%%%%%%%%%%%%%%%%%%%%%%%%%%%%
%%%%%%%%%%%%%%%%%%%%%%%%%%%%%%%%%%%%%%%%%%%%%%%%%%%%%%%%%%%%%%%%%%%%%%%%%%%%
%%%%%%%%%%%%%%%%%%%%%%%%%%%%%%%%%%%%%%%%%%%%%%%%%%%%%%%%%%%%%%%%%%%%%%%%%%%%
%%%%%%%%%%%%%%%%%%%%%%%%%%%%%%%%%%%%%%%%%%%%%%%%%%%%%%%%%%%%%%%%%%%%%%%%%%%%
%%%%%%%%%%%%%%%%%%%%%%%%%%%%%%%%%%%%%%%%%%%%%%%%%%%%%%%%%%%%%%%%%%%%%%%%%%%%
%%%%%%%%%%%%%%%%%%%%%%%%%%%%%%%%%%%%%%%%%%%%%%%%%%%%%%%%%%%%%%%%%%%%%%%%%%%%
%%%%%%%%%%%%%%%%%%%%%%%%%%%%%%%%%%%%%%%%%%%%%%%%%%%%%%%%%%%%%%%%%%%%%%%%%%%%
%%%%%%%%%%%%%%%%%%%%%%%%%%%%%%%%%%%%%%%%%%%%%%%%%%%%%%%%%%%%%%%%%%%%%%%%%%%%
%%%%%%%%%%%%%%%%%%%%%%%%%%%%%%%%%%%%%%%%%%%%%%%%%%%%%%%%%%%%%%%%%%%%%%%%%%%%
%%%%%%%%%%%%%%%%%%%%%%%%%%%%%%%%%%%%%%%%%%%%%%%%%%%%%%%%%%%%%%%%%%%%%%%%%%%%
%%%%%%%%%%%%%%%%%%%%%%%%%%%%%%%%%%%%%%%%%%%%%%%%%%%%%%%%%%%%%%%%%%%%%%%%%%%%
%%%%%%%%%%%%%%%%%%%%%%%%%%%%%%%%%%%%%%%%%%%%%%%%%%%%%%%%%%%%%%%%%%%%%%%%%%%%

\section{The low-temperature limit}
\label{lowTlimit}

At very low temperatures, Fermi-liquid theory applies. Then, the
Green's function (\ref{Green}) can be approximated by expanding the
self-energy to second order in $\omega$ thus capturing finite-lifetime
effects of the quasiparticles:
\begin{equation}
 \Sigma(\omega)=(1-1/Z)\omega+\alpha\omega^2+i\gamma(\omega).
\label{self-energy}
\end{equation}
Here,
$Z=(1-{\partial\Re\Sigma(\omega)/\partial\omega}|_{\omega=0})^{-1}$ is
the quasiparticle residue at the Fermi level and 
$\alpha=(1/2){\partial^2\Re\Sigma(\omega)/\partial\omega^2}|_{\omega=0}$.
Moreover,
\begin{equation}
 \gamma(\omega)=\tilde{\gamma}[T^2+({\omega/\pi})^2]
	\equiv\gamma_0(T)[1+({\beta\omega/\pi})^2]
\label{damping}
\end{equation}
is the quasiparticle damping. Due to Eq.\ (\ref{self-energy}), the
spectral function becomes
\begin{equation}
 A(\omega,\epsilon)=L_{\gamma(\omega)}({\omega\over Z}-\alpha\omega^2+
		\tilde{\mu}-\epsilon),
\label{spectral2}
\end{equation}
where $L_{\Gamma}(\omega)={1\over\pi}{\Gamma\over\Gamma^2+\omega^2}$
is the Lorentzian normalized to unity and
$\tilde{\mu}=\mu-\Re\Sigma(0)$ is the effective chemical potential.
In the limits $\omega\rightarrow0$ and $T\rightarrow0$, Eq.\
(\ref{spectral2}) reduces to the correct result,
$\lim_{\omega\rightarrow0}\lim_{T\rightarrow0}
A(\omega,\epsilon)=\delta(\epsilon-\tilde{\mu})$
\cite{Mueller-Hartmann:1989b}. 
Using also the sum rules $\int_{-\infty}^{\infty}d\epsilon\,\left[
L_{\Gamma}(\epsilon)\right]^{n+1}=(2n-1)!!/(2\pi\Gamma)^nn!$ for integer
$n$ and the fact that the width of $L_{\Gamma}(\epsilon)$ is
$1/\tau=2\Gamma$, we can check that Eqs.\ (\ref{condxx})-(\ref{condM})
reduce to the standard Boltzmann-theory results as $T\rightarrow0$. 
Since $A(\omega,\epsilon)$ of Eq.\ (\ref{spectral2}) is only the
low-frequency part of the spectral function, it only carries a small
fraction of the total spectral weight determined by the sum rule. 
When calculating transport coefficients, this does not matter:
Incoherent contributions missing in Eq.\ (\ref{spectral2}) are cut off
by $-{\partial f(\omega)/\partial\omega}$. 

Before proceeding, we investigate how the quasiparticle damping
depends on the correlation strength. From Eqs.\ (\ref{DOS}),
(\ref{localspectral}) and (\ref{spectral2}), we obtain the width of
the Abrikosov-Suhl resonance: 
\begin{equation}
\label{Tstar}
 T^*\simeq ZD.
\end{equation}
$T^*$ is a measure for the Fermi energy of the quasiparticles. At zero
temperature, $T^*$ is the only energy scale in the problem. Then, the 
self-energy is given by $\Sigma(\omega)=a\omega/T^*+ib(\omega/T^*)^2$
for $\omega\ll T^*$ with some constants $a$ and $b$. By comparing this
expression to the Fermi-liquid expansion (\ref{self-energy}), we find
$1/T^*\propto1-1/Z$ and $\tilde{\gamma}\propto1/{T^*}^2$. The former
relation is consistent with Eq.\ (\ref{Tstar}) since in the
vicinity of the density-driven Mott transition, $Z\sim\delta$
\cite{Kajueter:1996a}. By inserting the other relation,
$\tilde{\gamma}\propto1/{T^*}^2$, into the definition of $\gamma_0$ in
Eq.\ (\ref{damping}) and by also using Eq.\ (\ref{Tstar}), we obtain
\begin{equation}
 \gamma_0=A\,\frac{T^2}{Z^2D},
\label{damping2}
\end{equation}
where $A$ is a dimensionless number and which is valid close to the
density-driven Mott transition.

Using Eqs.\ (\ref{damping}) and (\ref{spectral2}), we find that to
leading order in $T$, the conductivities (\ref{condxx})-(\ref{condM})
do not depend on $\alpha$:
\begin{eqnarray}
 \sigma_{xx}&=&{N_se^2\over2\gamma_0}\;
	\phi_{xx}(\tilde{\mu})E^{(1)}\label{con2xx},\\
 \sigma_{xy}&=&\frac{N_se^3H}{4\gamma_0^2}
	\phi_{xy}(\tilde{\mu})E^{(2)}\label{con2xy},\\
 \Delta\sigma_{xx}&=&\frac{N_se^4H^2}{8\gamma_0^3}\;
        \phi_{M}(\tilde{\mu})E^{(3)}\label{con2M},
\end{eqnarray}
where the numbers $E^{(l)}$ are given by 
\begin{equation}
 E^{(l)}=\int_{-\infty}^{\infty}{dx \over 4\cosh^{2}\left({x\over2}\right)
	\left[1+\left({x\over\pi}\right)^{2}\right]^l}.
\label{Enumbers}
\end{equation}
Numerically, we obtain $E^{(1)}=0.822467$, $E^{(2)}=0.711748$, and
$E^{(3)}=0.635279$. Using Eqs.\ (\ref{resiH}), (\ref{resiM}), and
(\ref{con2xx})-(\ref{con2M}), we find
\begin{eqnarray}
 R_H&=&{1\over N_se}\;{\phi_{xy}(\tilde{\mu})E^{(2)}\over\left[
	\phi_{xx}(\tilde{\mu})E^{(1)}\right]^2}\label{res2H},\\
 \Delta\rho&=&{H^2\over2N_s\gamma_0}\;
	{\left|\phi_M(\tilde{\mu})\right|E^{(3)}\over\left[
	\phi_{xx}(\tilde{\mu})E^{(1)}\right]^2}\label{res2M}.
\end{eqnarray} 
As already mentioned, the coupling of the magnetic field to the
orbital motion of the electrons gives rise to a positive contribution
to the resistivity. 

The Hall constant (\ref{res2H}) does not depend on the quasiparticle
damping. However, the ordinary resistivity and the magnetoresistance
do. Near the density-driven Mott transition, these quantities are
given, due to Eq.\ (\ref{damping2}), by
\begin{eqnarray}
 \rho&=&\frac{2AT^2}{N_se^2Z(\tilde{\mu})^2D\phi_{xx}(\tilde{\mu})
	E^{(1)}},\label{res3}\\
 \Delta\rho&=&{H^2Z(\tilde{\mu})^2D\over2N_sA\,T^2}\;
	{\left|\phi_M(\tilde{\mu})\right|E^{(3)}\over\left[
	\phi_{xx}(\tilde{\mu})E^{(1)}\right]^2}\label{res3M}.
\end{eqnarray} 
The impact of correlations on both quantities is described by their
dependence on $Z$: While they enhance the ordinary resistivity by a
factor of $1/Z^2$, they lower the magnetoresistance by $Z^2$, in each
case relative to a noninteracting system with the same density of
states. Also, the dependences of the quantities (\ref{res3}) and
(\ref{res3M}) on the doping level are mainly determined by $Z$: As
the Mott transition is approached, $\delta\rightarrow0$, $\rho$
diverges and the magnetoresistance vanishes. The former fact indicates
that the effective charge carriers become localized. But then, the
magnetic field can no longer affect the orbital motion of the
electrons, hence the vanishing magnetoresistance. 

Finally, the Fermi-liquid relations $\rho\propto T^2/Z^2$ and
$\Delta\rho\propto Z^2/T^2$ imply the validity of Kohler's rule
\cite{Ziman1}, 
\begin{equation}
 {\Delta\rho\over\rho}\propto\left({H\over\rho}\right)^2.
\label{Kohler}
\end{equation}
 
%%%%%%%%%%%%%%%%%%%%%%%%%%%%%%%%%%%%%%%%%%%%%%%%%%%%%%%%%%%%%%%%%%%%%%%%%%%%%
%%%%%%%%%%%%%%%%%%%%%%%%%%%%%%%%%%%%%%%%%%%%%%%%%%%%%%%%%%%%%%%%%%%%%%%%%%%%%
%%%%%%%%%%%%%%%%%%%%%%%%%%%%%%%%%%%%%%%%%%%%%%%%%%%%%%%%%%%%%%%%%%%%%%%%%%%%%
%%%%%%%%%%%%%%%%%%%%%%%%%%%%%%%%%%%%%%%%%%%%%%%%%%%%%%%%%%%%%%%%%%%%%%%%%%%%%
%%%%%%%%%%%%%%%%%%%%%%%%%%%%%%%%%%%%%%%%%%%%%%%%%%%%%%%%%%%%%%%%%%%%%%%%%%%%%
%%%%%%%%%%%%%%%%%%%%%%%%%%%%%%%%%%%%%%%%%%%%%%%%%%%%%%%%%%%%%%%%%%%%%%%%%%%%%
%%%%%%%%%%%%%%%%%%%%%%%%%%%%%%%%%%%%%%%%%%%%%%%%%%%%%%%%%%%%%%%%%%%%%%%%%%%%%
%%%%%%%%%%%%%%%%%%%%%%%%%%%%%%%%%%%%%%%%%%%%%%%%%%%%%%%%%%%%%%%%%%%%%%%%%%%%%
%%%%%%%%%%%%%%%%%%%%%%%%%%%%%%%%%%%%%%%%%%%%%%%%%%%%%%%%%%%%%%%%%%%%%%%%%%%%%
%%%%%%%%%%%%%%%%%%%%%%%%%%%%%%%%%%%%%%%%%%%%%%%%%%%%%%%%%%%%%%%%%%%%%%%%%%%%%
%%%%%%%%%%%%%%%%%%%%%%%%%%%%%%%%%%%%%%%%%%%%%%%%%%%%%%%%%%%%%%%%%%%%%%%%%%%%%
%%%%%%%%%%%%%%%%%%%%%%%%%%%%%%%%%%%%%%%%%%%%%%%%%%%%%%%%%%%%%%%%%%%%%%%%%%%%%
%%%%%%%%%%%%%%%%%%%%%%%%%%%%%%%%%%%%%%%%%%%%%%%%%%%%%%%%%%%%%%%%%%%%%%%%%%%%%
%%%%%%%%%%%%%%%%%%%%%%%%%%%%%%%%%%%%%%%%%%%%%%%%%%%%%%%%%%%%%%%%%%%%%%%%%%%%%
%%%%%%%%%%%%%%%%%%%%%%%%%%%%%%%%%%%%%%%%%%%%%%%%%%%%%%%%%%%%%%%%%%%%%%%%%%%%%
%%%%%%%%%%%%%%%%%%%%%%%%%%%%%%%%%%%%%%%%%%%%%%%%%%%%%%%%%%%%%%%%%%%%%%%%%%%%%

\section{The high-temperature limit}
\label{highTlimit}  

To single out the contributions of the incoherent parts of the spectrum,
we now consider the limit of very high temperatures. For the Hubbard
interaction to retain its important impact on the electron dynamics in
the high-temperature limit, we have to choose $U=\infty$ from the
outset. Then, the spectral function (\ref{spectral}) does no longer
receive contributions from the upper Hubbard band. We start by
deriving a sum rule for the local spectral function
(\ref{localspectral}). On average, each site hosts $n$ electrons and
$1-n$ holes, since $U=\infty$ means that the maximum occupancy is
1. Thus, we find
\begin{eqnarray}
 \int_{-\infty}^{\infty}d\omega\,A(\omega)f(\omega)&=&n/N_s
\label{defchem},\\
 \int_{-\infty}^{\infty}d\omega\,A(\omega)[1-f(\omega)]&=&1-n
\label{towardssumrule}.
\end{eqnarray} 
These equations imply the sum rule
\begin{equation}
 \int_{-\infty}^{\infty}d\omega\,A(\omega)=1-n+n/N_s.
\label{sumrule}
\end{equation}

Next, we derive the high-temperature expansion for the chemical
potential, which is defined by Eq.\ (\ref{defchem}). In the limit
$T\rightarrow\infty$, the spectral function (\ref{localspectral})
becomes temperature independent except for the presumably
temperature-dependent peak center, $-\mu(T)$. Therefore, we introduce
tilded spectral and Green's functions that are obtained from their
untilded counterparts by replacing $\omega\rightarrow\omega-\mu$. For
instance,
\begin{equation}
 \tilde{A}(\omega)={A}(\omega-\mu).
\label{tildelocalspectral}
\end{equation}
Eq.\ (\ref{defchem}) can now be rewritten as
$\int_{-\infty}^{\infty}d\omega\,\tilde{A}(\omega)f(\omega-\mu)=n/N_s$.
By expanding $f(\omega-\mu)$ about the peak center of
$\tilde{A}(\omega)$, $\omega=0$, by further expressing all derivatives
of the Fermi function by the Fermi function itself, and by using the
moments 
\begin{equation}
 m_l=\int_{-\infty}^{\infty}d\omega\,\tilde{A}(\omega)\omega^l
\label{momentsofspectral}
\end{equation}
along with their expansion in powers of $1/T$,
\begin{equation}
 m_l=\sum_{n=0}^{\infty}\beta^nm_l^{(n)},
\label{momentsexpansion}
\end{equation}
we can systematically expand $f(-\mu(T))$ in powers of $1/T$. To first
order in $1/T$, we find
\begin{equation}
 f(-\mu(T))={n\over N_s}\left[{1\over
   m_0}+\beta\,\delta\frac{m_1^{(0)}}{m_0^3}\right]+o(\beta^2), 
\label{chemexpansion}
\end{equation}
where $m_0=1-n+n/N_s$ according to Eq.\ (\ref{sumrule}).

Likewise, we can expand the conductivities (\ref{condxx})-(\ref{condM})
in powers of $1/T$. In terms of the quantities
\begin{eqnarray}
 C_l&=&\int d\epsilon\,\phi_{xx}(\epsilon)
 	\int d\omega\,\tilde{A}(\omega,\epsilon)^2\omega^l,\label{gammxx}\\
 C_l^H&=&\int d\epsilon\,\phi_{xy}(\epsilon)
 	\int d\omega\,\tilde{A}(\omega,\epsilon)^3\omega^l,\label{gammxy}\\
 C_l^M&=&\int d\epsilon\,\phi_{M}(\epsilon)
 	\int d\omega\,\tilde{A}(\omega,\epsilon)^4\omega^l,\label{gammM}
\end{eqnarray}
which are to be calculated in the high-temperature limit, we find
\begin{eqnarray}
 \sigma_{xx}&=&N_s\pi e^2\beta A(1-A)C_0,\label{hiTcondxx}\\
 \sigma_{xy}&=&\frac{N_s 2\pi^2e^3H}{3}\beta A(1-A)
	[C_0^H-\beta(1-2A)C_1^H], 
	\label{hiTcondxy}\\
 \Delta\sigma_{xx}&=&\frac{N_s2\pi^3e^4H^2}{5}\beta A(1-A)C_0^M
	\label{hiTcondM},
\end{eqnarray}
where $A\equiv\frac{n/N_s}{1-n+n/N_s}$. Using Eqs.\ (\ref{resiH})
and (\ref{resiM}), these results can be translated into resistivities,
\begin{eqnarray}
 \rho&=&\frac{T}{\pi e^2N_s\zeta C_0},\label{hiTresxx}\\
 R_H&=&\frac{2\left(TC_0^H-\eta\,C_1^H\right)}
	{3eN_s\zeta C_0^2},\label{hiTresxy}\\
 \Delta\rho&=&\frac{2\pi H^2\left|C_0^M\right|}{5N_s\zeta C_0^2}\;T,
	\label{hiTresM}
\end{eqnarray}
where we have defined $\zeta\equiv A(1-A)$ and $\eta\equiv1-2A$, or
\begin{eqnarray}
 \zeta&=&\frac{(1-n)n/N_s}{(1-n+n/N_s)^2}\label{zeta},\\
 \eta&=&\frac{1-n-n/N_s}{1-n+n/N_s}\label{eta}.
\end{eqnarray}

Before discussing these findings, we investigate the dependences of
the coefficients (\ref{gammxx})-(\ref{gammM}) on the band filling and
on temperature. Their doping dependence can roughly be estimated by
assuming that the weight of the full spectral function, Eq.\
(\ref{spectral}), is similar to that of the local one in Eq.\
(\ref{localspectral}). This implies
\begin{eqnarray}
 C_l&\approx&\tilde{C}_l(1-n+n/N_s)^2,\label{dopgammxx}\\
 C_l^H&\approx&\tilde{C}_l^H(1-n+n/N_s)^3,\label{dopgammxy}\\
 C_l^M&\approx&\tilde{C}_l^M(1-n+n/N_s)^4,\label{dopgammM}
\end{eqnarray}
where $\tilde{C}_l$, $\tilde{C}_l^H$, and
$\tilde{C}_l^M$ are doping independent. Eqs.\
(\ref{dopgammxx})-(\ref{dopgammM}) can be shown to be a good
approximation in the vicinity of half filling. 

The temperature dependence of the coefficients
(\ref{gammxx})-(\ref{gammM}) is more subtle because it can be affected
by symmetry. $C_0$ and $C_0^M$ are positive and negative definite,
respectively, and therefore tend to constants determined by
the limiting form of the spectral function as $T\rightarrow\infty$. 
However, $C_0^H$ receives contributions of either sign leading to
the possibility that these contributions may cancel each other out
in leading order due to some symmetry. As an example, we will
demonstrate further down that perfect nesting leads to
$C_0^H\sim1/T$. Then, both contributions in Eq.\ (\ref{hiTresxy})
matter and the Hall constant tends to a constant as
$T\rightarrow\infty$. For a general band structure, however, there is
no reason why $C_0^H$ should vanish as $T\rightarrow\infty$, so
$\sigma_{xy}\sim 1/T$ like the longitudinal conductivity, and hence,
$R_H\sim T$ in this limit. In the case of the simple tight-binding
band of Eq.\ (\ref{dispersion}), we evaluate the coefficients
appearing in Eqs.\ (\ref{hiTresxx})-(\ref{hiTresM}) numerically within
the NCA for $\delta=0.1$: Setting $d=3$, we find
$\lim_{T\rightarrow\infty}C_0=1.894\times10^{-2}$,
$\lim_{T\rightarrow\infty}TC_0^H=-2.654\times10^{-4}$,
$\lim_{T\rightarrow\infty}C_1^H=-7.602\times10^{-4}$, and 
$\lim_{T\rightarrow\infty}C_0^M=-4.529\times10^{-4}$.

We now demonstrate that any band structure implying the
symmetry properties
\begin{eqnarray}
 D(-\epsilon)&=&D(\epsilon)\label{DOSsym},\\
 \phi_{xy}(-\epsilon)&=&-\phi_{xy}(\epsilon)\label{phixysym}
\end{eqnarray}
in fact leads to $C_0^H\sim1/T$. These properties can be
satisfied, for instance, on account of the perfect-nesting condition
\begin{equation}
 \epsilon(\vec{k}+\vec{Q})=-\epsilon(\vec{k})
\label{nesting}
\end{equation}
for some vector $\vec{Q}$. In the case of nearest-neighbor hopping on
a cubic lattice, Eq.\ (\ref{dispersion}), $\vec{Q}$ is the vector
pointing to the corner of the Brillouin zone, i.e.,
$\vec{Q}=(\pi,\pi,...)$. Upon solving the dynamical mean-field
equations, an even density of states as in Eq.\ (\ref{DOSsym}) is seen
to lead to an even local spectral function as $T\rightarrow\infty$:
\begin{equation}
 \lim_{T\rightarrow\infty}\tilde{A}(-\omega)=
 \lim_{T\rightarrow\infty}\tilde{A}(\omega).
\label{localspectralsym}
\end{equation}
Finally, by again using the dynamical mean-field equations, we can
show that Eq.\ (\ref{localspectralsym}) implies a symmetry property
for the full spectral function: 
\begin{equation}
 \lim_{T\rightarrow\infty}\tilde{A}(-\omega,-\epsilon)=
 \lim_{T\rightarrow\infty}\tilde{A}(\omega,\epsilon).
\label{fullspectralsym}
\end{equation}
Eqs.\ (\ref{phixysym}) and (\ref{fullspectralsym}) prove that
$\lim_{T\rightarrow\infty}C_0^H=0$. The first order in $1/T$ does
not vanish, hence $C_0^H\sim1/T$ as claimed above.

We now discuss our results for the three quantities in Eqs.\
(\ref{hiTresxx})-(\ref{hiTresM}). Both the ordinary resistivity and
the magnetoresistance exhibit a linear-in $T$ behavior at high
temperatures with prefactors that diverge in the empty-band limit or
as half filling is approached. As $n\rightarrow0$, we find
$\rho,\Delta\rho\sim{T/n}$, which simply reflects the fact that if
the charge carriers vanish, the conductivity has to vanish too. Close
to half filling, we find
\begin{eqnarray}
 \rho&\sim&{T/\delta},\label{resqual}\\
 \Delta\rho&\sim&{T/N_s^2\delta}.\label{resMqual}
\end{eqnarray}
At the Mott transition, these quantities diverge because localized
charge carriers cannot give rise to a finite conductivity either.

The temperature dependence of the Hall constant depends on whether the
given band structure satisfies the conditions (\ref{DOSsym}) and
(\ref{phixysym}). If they are satisfied, $R_H\sim\mbox{const}$,
otherwise, we expect $R_H\sim T$. This result is consistent with
previous works on the high-temperature Hall effect by 
Brinkman and Rice \cite{Brinkman:1971}, Oguri and
Maekawa \cite{Oguri:1990}, and Shastry, Shraiman, and
Singh \cite{Shastry:1993}. The high-temperature expansions of all
these works were carried out for a cubic lattice with nearest-neighbor
hopping. Therefore, they all obtained $\sigma_{xy}\sim1/T^2$, leading
to a temperature-independent Hall constant at high temperatures. Our
analysis shows that this result is not generic. Rather, we expect
$\sigma_{xy}\sim1/T$ and thus $R_H\sim T$ for a general band structure
at high temperatures. To check this, we calculate the
infinite-frequency Hall constant $R_H^*$ of Ref.\ \cite{Shastry:1993}
to leading order in $1/T$ on a two-dimensional cubic lattice where
electrons can hop with amplitudes $t$ and $t^{(d)}$ to, respectively, 
nearest-neighbor sites and diagonally across the unit cell. In
essence, $R_H^*$ is given by $\langle[\hat{J}_x,\hat{J}_y]\rangle$. 
For this quantity to be of order $1/T$ rather than $1/T^2$, electrons
must be able to circumscribe a finite area enclosing a finite flux
with just three hops. This is made possible by the inclusion of
diagonal hops $t^{(d)}$. To leading order in $1/T$
\cite{Thompson:1991}, we find
$R_H^*={6t^2t^{(d)}\over(t^2+2{t^{(d)}}^2)^2}  
{(1+\delta)T\over e\delta(1-\delta)}$. A high-temperature behavior
$R_H\sim T$ is also striking from the viewpoint of conventional band
theory, where a temperature dependence can only arise below a scale
set by the Debye temperature.

As for the doping dependence of the Hall constant, we find either
$R_H\sim1/N_s\delta|e|$ or $R_H\sim T/N_s\delta|e|$ close to half
filling. In any case, the Hall constant exhibits a $1/\delta$
behavior. The sign of the Hall constant is governed by the
coefficients $C_0^H$ and $C_1^H$. As can be seen from Eq.\
(\ref{gammxy}), they are a combined result of the band structure,
entering via the function (\ref{phixy}), and the correlations, taken
into account by the spectral function. In the case of the simple
tight-binding band (\ref{dispersion}), the Hall constant is positive
in the vicinity of half filling.

Finally, Eqs.\ (\ref{resqual}) and (\ref{resMqual}) indicate that
Kohler's rule is replaced by $\Delta\rho/\rho\sim\mbox{const}\times
H^2$ in the high-temperature limit. Moreover, the Hall angle, defined
by $\cot\theta_H=\rho/R_H$, increases at most as $\cot\theta_H\sim T$
as a function of temperature, if the symmetry properties
(\ref{DOSsym}) and (\ref{phixysym}) are satisfied. Otherwise,
$\cot\theta_H\sim \mbox{const}$.  
 
%%%%%%%%%%%%%%%%%%%%%%%%%%%%%%%%%%%%%%%%%%%%%%%%%%%%%%%%%%%%%%%%%%%%%%%%%%%%%
%%%%%%%%%%%%%%%%%%%%%%%%%%%%%%%%%%%%%%%%%%%%%%%%%%%%%%%%%%%%%%%%%%%%%%%%%%%%%
%%%%%%%%%%%%%%%%%%%%%%%%%%%%%%%%%%%%%%%%%%%%%%%%%%%%%%%%%%%%%%%%%%%%%%%%%%%%%
%%%%%%%%%%%%%%%%%%%%%%%%%%%%%%%%%%%%%%%%%%%%%%%%%%%%%%%%%%%%%%%%%%%%%%%%%%%%%
%%%%%%%%%%%%%%%%%%%%%%%%%%%%%%%%%%%%%%%%%%%%%%%%%%%%%%%%%%%%%%%%%%%%%%%%%%%%%
%%%%%%%%%%%%%%%%%%%%%%%%%%%%%%%%%%%%%%%%%%%%%%%%%%%%%%%%%%%%%%%%%%%%%%%%%%%%%
%%%%%%%%%%%%%%%%%%%%%%%%%%%%%%%%%%%%%%%%%%%%%%%%%%%%%%%%%%%%%%%%%%%%%%%%%%%%%
%%%%%%%%%%%%%%%%%%%%%%%%%%%%%%%%%%%%%%%%%%%%%%%%%%%%%%%%%%%%%%%%%%%%%%%%%%%%%
%%%%%%%%%%%%%%%%%%%%%%%%%%%%%%%%%%%%%%%%%%%%%%%%%%%%%%%%%%%%%%%%%%%%%%%%%%%%%
%%%%%%%%%%%%%%%%%%%%%%%%%%%%%%%%%%%%%%%%%%%%%%%%%%%%%%%%%%%%%%%%%%%%%%%%%%%%%
%%%%%%%%%%%%%%%%%%%%%%%%%%%%%%%%%%%%%%%%%%%%%%%%%%%%%%%%%%%%%%%%%%%%%%%%%%%%%
%%%%%%%%%%%%%%%%%%%%%%%%%%%%%%%%%%%%%%%%%%%%%%%%%%%%%%%%%%%%%%%%%%%%%%%%%%%%%
%%%%%%%%%%%%%%%%%%%%%%%%%%%%%%%%%%%%%%%%%%%%%%%%%%%%%%%%%%%%%%%%%%%%%%%%%%%%%
%%%%%%%%%%%%%%%%%%%%%%%%%%%%%%%%%%%%%%%%%%%%%%%%%%%%%%%%%%%%%%%%%%%%%%%%%%%%%
%%%%%%%%%%%%%%%%%%%%%%%%%%%%%%%%%%%%%%%%%%%%%%%%%%%%%%%%%%%%%%%%%%%%%%%%%%%%%
%%%%%%%%%%%%%%%%%%%%%%%%%%%%%%%%%%%%%%%%%%%%%%%%%%%%%%%%%%%%%%%%%%%%%%%%%%%%%

\section{Intermediate temperatures}
\label{intTregime}

The behavior of the Hall coefficient at intermediate temperatures
and for some values of the parameter $U$ have been studied numerically
in Refs.\ \cite{Pruschke:1995} and \cite{Majumdar:1995}. In these
works, it was stressed that the Hubbard model in infinite
dimensions reproduces many observable features of the high-temperature
superconductors. Other works have shown, however, that important
quantities of these systems such as the specific heat are not
properly described within the infinite-$d$ Hubbard model and that this
model is more appropriate for describing some transition-metal oxides
where due to the orbital degeneracy, short-ranged magnetic
correlations are weaker than in the cuprates. Nevertheless, given that
we do not yet understand the physics of the cuprates, we feel that the
observation of similarities between experimental results on the Hall
effect in cuprates and numerical results for the infinite-$d$ Hubbard
model deserves further investigation. The magnetoresistance is a
second probe of the effects of a magnetic field. In this section, we
carry out a comparative study of these two quantities with the goal of
elucidating which elements of the physics of the large-$d$ Hubbard
model result in the observed similarities with the cuprates. To this
effect, we restrict the discussion to the simple band structure
(\ref{dispersion}).

\begin{center}
\begin{figure}
	\epsfig{file=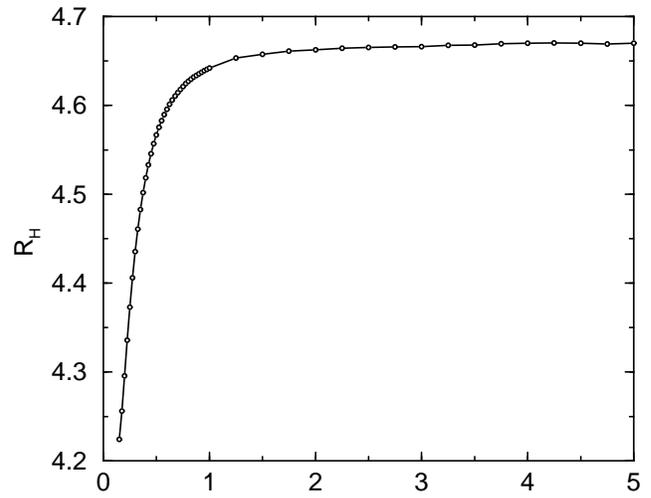,%
	width=2.6in,angle=-90,bbllx=93pt,bblly=49pt,bburx=564pt,bbury=636pt}
	\vspace{6pt}
	\caption{Hall constant at $U=\infty$ and $\delta=0.1$.}
\label{fig:hIconst}
\end{figure}
\end{center}

\begin{center}
\begin{figure}
	\epsfig{file=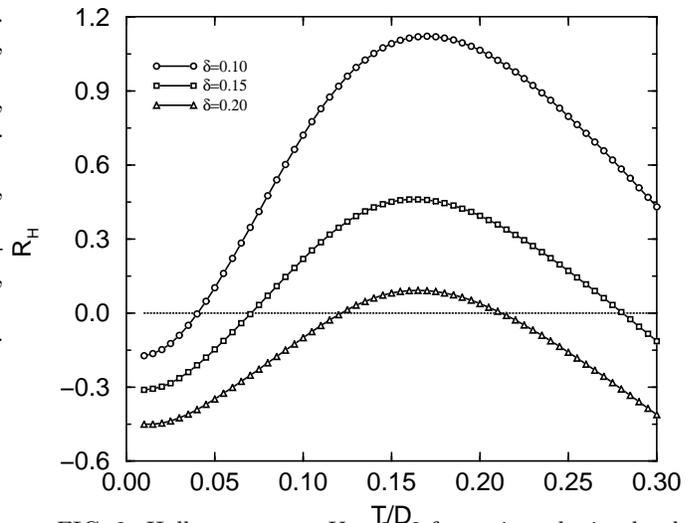,%
	width=2.6in,angle=-90,bbllx=93pt,bblly=49pt,bburx=564pt,bbury=636pt}
	\vspace{6pt}
	\caption{Hall constant at $U=2.82$ for various doping levels.}
\label{fig:h2const}
\end{figure}
\end{center}

\begin{center}
\begin{figure}
	\epsfig{file=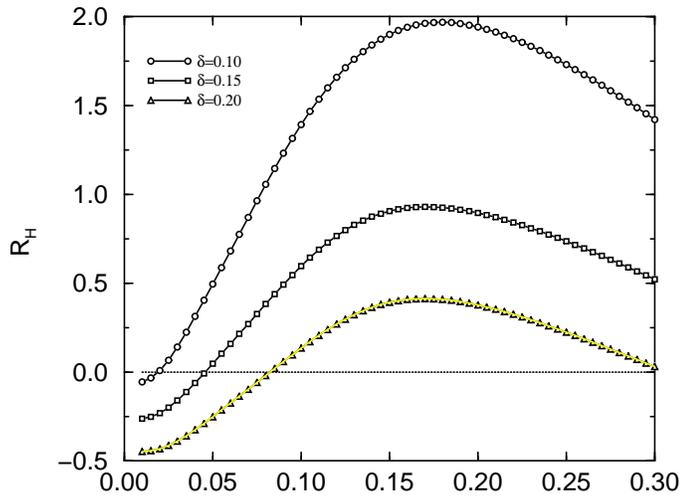,%
	width=2.6in,angle=-90,bbllx=93pt,bblly=49pt,bburx=564pt,bbury=636pt}
	\vspace{6pt}
	\caption{Hall constant at $U=4$ for various doping levels.}
\label{fig:h4const}
\end{figure}
\end{center}

The Hall constant is shown in Figs.\
\ref{fig:hIconst}-\ref{fig:h4const}: We know from Sec.\ 
\ref{lowTlimit} that the Hall constant starts at its noninteracting
value at $T=0$. We also know that within the infinite-$U$ Hubbard
model, it rapidly converges to a positive value beyond the high-energy
scale $D$ (cf.\ Fig.\ \ref{fig:hIconst}). Moreover, this positive
value goes like $1/\delta$ in the vicinity of half filling. In between
these limiting cases, we 
expect a smooth crossover implying a sign change of the Hall constant
as a function of temperature. At finite $U$, however, the Hall constant
goes through a maximum as a function of temperature (Figs.\
\ref{fig:h2const} and \ref{fig:h4const}). For big enough $U$ and small
enough doping levels, this maximum is positive. For temperatures
greater than the Mott-Hubbard gap, the dynamics of the electrons is
increasingly insensitive to the interaction $U$ which is why the Hall
constant becomes electron-like again at very high temperatures. The
position of the maximum depends only weakly on the doping level and is
roughly located at $T\approx0.16D$, which is above $T^*$ (Figs.\
\ref{fig:h2const} and \ref{fig:h4const}). Upon increasing $U$, the
maximum becomes little by little more asymmetric and ultimately
approaches the $U=\infty$ form of the Hall constant discussed
above. This indicates that the decrease of the Hall constant as a
function of temperature beyond its maximum is due to the excitation of
charge across the Mott-Hubbard gap. This does not imply that the
relevant temperature scale is $U-2D$, since in dynamical 
mean-field theory, the two Hubbard bands are only separated by a
pseudogap \cite{Georges:1993b,Kajueter:1996a}. In fact, Figs.\
\ref{fig:h2const} and \ref{fig:h4const} show that the relevant
temperature scale of the decrease is roughly $0.1D$. At first glance,
the curves of Figs.\ \ref{fig:h2const} and \ref{fig:h4const} bear some
similarities to what is found experimentally in cuprates
\cite{Ong:1990,Hwang:1994}. In cuprates, the width of the lower Cu
band is roughly $0.5$ eV implying $0.1D\sim250$ K. This scale does in
fact roughly characterize the experimentally observed decrease
\cite{Hwang:1994}. Yet, despite these similarities, there is a crucial
difference:  In cuprates, the two relevant bands are separated by a
{\it real} gap of approximately 3 eV, which is about six times as
large as the width of the lower Cu band. By contrast, in Fig.\
\ref{fig:h4const}, these bands are separated by only a 
{\it pseudogap} which is only about as big as the width of the lower
Hubbard band, namely, $U-2D=2D$. Therefore, the situation in cuprates
is much closer to $U=\infty$ in our model. Given that the maximum in
the Hall constant found in Ref.\ \cite{Pruschke:1995} is a finite-$U$
effect, we conclude that the dynamical mean-field theory cannot
describe the experimentally observed decrease of the Hall constant as
a function of temperature unless a reason is found to introduce a much
smaller Hubbard gap within the framework of the large-$d$ Hubbard
model than what is really observed experimentally.

\begin{center}
\begin{figure}
	\epsfig{file=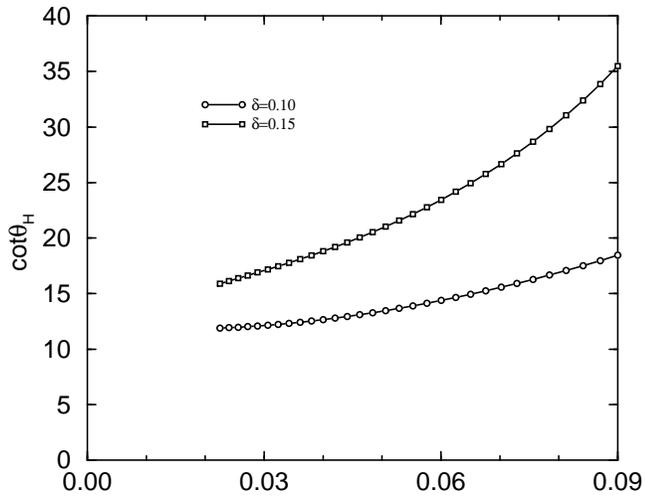,%
	width=2.6in,angle=-90,bbllx=93pt,bblly=49pt,bburx=564pt,bbury=636pt}
	\vspace{6pt}
	\caption{Hall angle at $U=4$ for two doping levels.}
\label{fig:h4angle}
\end{figure}
\end{center}

\begin{center}
\begin{figure}
	\epsfig{file=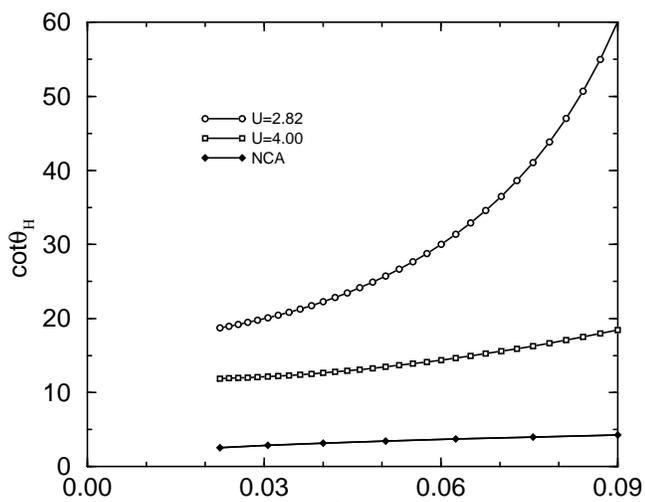,%
	width=2.6in,angle=-90,bbllx=93pt,bblly=49pt,bburx=564pt,bbury=636pt}
	\vspace{6pt}
	\caption{Hall angle for various $U$ and $\delta=0.1$. The NCA
	curve corresponds to $U=\infty$.}
\label{fig:hUangle}
\end{figure}
\end{center}

We now turn to the Hall angle. From our analytical investigation, we
know that $\cot\theta_H\sim T$ for $T>D$ and $\cot\theta_H\sim T^2$
for $T<T_{\mbox{\small coh}}$. In cuprates such as
La$_{2-x}$Sr$_x$CuO$_4$, a quadratic temperature dependence of the
Hall angle is observed in the underdoped compounds \cite{Chien:1991},
where at the same time the Hall constant is holelike. 
In Figs.\ \ref{fig:h4angle} and \ref{fig:hUangle}, the Hall
angle is shown as a function of $T^2$ in the intermediate-temperature
regime, $0.15D\le T\le0.3D$, where the Hall constant is
positive. Fig.\ \ref{fig:h4angle} shows that $\cot\theta_H$ becomes
smaller upon decreasing the doping level, as seen in the experiment
\cite{Hwang:1994}. However, none of the curves displays linear
behavior although a linear fit becomes better as $U\rightarrow\infty$
and for sufficiently small temperatures and doping levels (Fig.\
\ref{fig:hUangle}). Note that a linear extrapolation of the $U=\infty$
curve in Fig.\ \ref{fig:hUangle} would lead to a finite intercept.

\begin{center}
\begin{figure}
	\epsfig{file=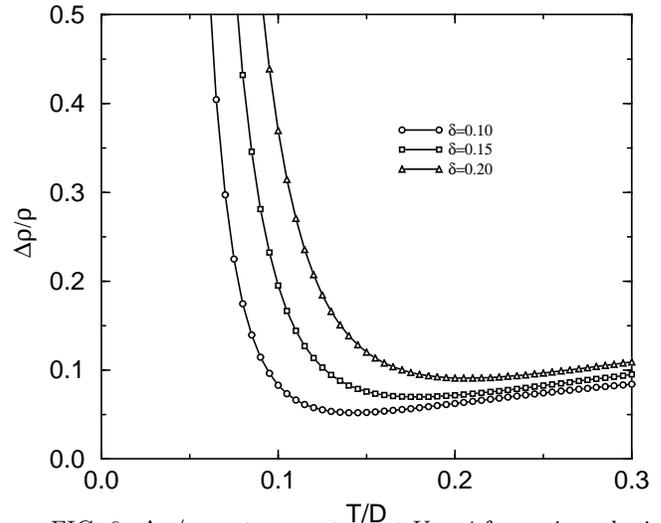,%
	width=2.6in,angle=-90,bbllx=93pt,bblly=49pt,bburx=564pt,bbury=636pt}
	\vspace{6pt}
	\caption{$\Delta\rho/\rho$ vs.\ temperature at $U=4$ for
		various doping levels.} 
\label{fig:m4resdres}
\end{figure}
\end{center}

\begin{center}
\begin{figure}
	\epsfig{file=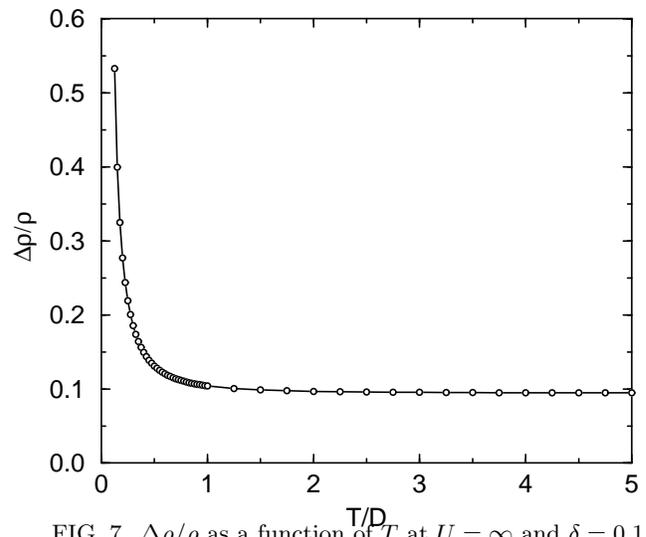,%
	width=2.6in,angle=-90,bbllx=93pt,bblly=49pt,bburx=564pt,bbury=636pt}
	\vspace{6pt}
	\caption{$\Delta\rho/\rho$ as a function of $T$ at $U=\infty$
    	 	and $\delta=0.1$.} 
\label{fig:mIresdres}
\end{figure}
\end{center}

Further evidence that the infinite-$d$ Hubbard model in the
intermediate-temperature regime is not able to explain the
magnetotransport observed in cuprates is provided by our study of the
magnetoresistance. In the Fermi-liquid regime, $T<T_{\mbox{\small
coh}}$, we obtained $\Delta\rho/\rho\sim1/T^4$, while in the opposite 
limit at very high temperatures, this ratio saturates to a
temperature-independent value. Fig.\ \ref{fig:m4resdres} displays
$\Delta\rho/\rho$ at $U=4$ for intermediate temperatures. 
Experimentally, a $1/T^4$ dependence is observed in the normal state
of high-purity cuprates \cite{Harris:1995}. By contrast, the curves of
Fig.\ \ref{fig:m4resdres} increase as a function of temperature in the
regime where the Hall constant decreases as a function of temperature. 
This fact is related to the appearance of a minimum whose location
depends on the doping level and which is a finite-$U$ effect too. In
fact, as can be seen in Fig.\ \ref{fig:mIresdres}, this minimum is
absent at $U=\infty$. There, in agreement with our analytical
prediction, the ratio $\Delta\rho/\rho$ approaches a constant value
beyond the high-energy scale $D$ and cannot, therefore, obey Kohler's
rule nor the modified version thereof suggested by Terasaki {\it et
al.} \cite{Terasaki:1995}. This is also true at intermediate
temperatures.

%%%%%%%%%%%%%%%%%%%%%%%%%%%%%%%%%%%%%%%%%%%%%%%%%%%%%%%%%%%%%%%%%%%%%%%%%%%%%
%%%%%%%%%%%%%%%%%%%%%%%%%%%%%%%%%%%%%%%%%%%%%%%%%%%%%%%%%%%%%%%%%%%%%%%%%%%%%
%%%%%%%%%%%%%%%%%%%%%%%%%%%%%%%%%%%%%%%%%%%%%%%%%%%%%%%%%%%%%%%%%%%%%%%%%%%%%
%%%%%%%%%%%%%%%%%%%%%%%%%%%%%%%%%%%%%%%%%%%%%%%%%%%%%%%%%%%%%%%%%%%%%%%%%%%%%
%%%%%%%%%%%%%%%%%%%%%%%%%%%%%%%%%%%%%%%%%%%%%%%%%%%%%%%%%%%%%%%%%%%%%%%%%%%%%
%%%%%%%%%%%%%%%%%%%%%%%%%%%%%%%%%%%%%%%%%%%%%%%%%%%%%%%%%%%%%%%%%%%%%%%%%%%%%
%%%%%%%%%%%%%%%%%%%%%%%%%%%%%%%%%%%%%%%%%%%%%%%%%%%%%%%%%%%%%%%%%%%%%%%%%%%%%
%%%%%%%%%%%%%%%%%%%%%%%%%%%%%%%%%%%%%%%%%%%%%%%%%%%%%%%%%%%%%%%%%%%%%%%%%%%%%
%%%%%%%%%%%%%%%%%%%%%%%%%%%%%%%%%%%%%%%%%%%%%%%%%%%%%%%%%%%%%%%%%%%%%%%%%%%%%
%%%%%%%%%%%%%%%%%%%%%%%%%%%%%%%%%%%%%%%%%%%%%%%%%%%%%%%%%%%%%%%%%%%%%%%%%%%%%
%%%%%%%%%%%%%%%%%%%%%%%%%%%%%%%%%%%%%%%%%%%%%%%%%%%%%%%%%%%%%%%%%%%%%%%%%%%%%
%%%%%%%%%%%%%%%%%%%%%%%%%%%%%%%%%%%%%%%%%%%%%%%%%%%%%%%%%%%%%%%%%%%%%%%%%%%%%
%%%%%%%%%%%%%%%%%%%%%%%%%%%%%%%%%%%%%%%%%%%%%%%%%%%%%%%%%%%%%%%%%%%%%%%%%%%%%
%%%%%%%%%%%%%%%%%%%%%%%%%%%%%%%%%%%%%%%%%%%%%%%%%%%%%%%%%%%%%%%%%%%%%%%%%%%%%
%%%%%%%%%%%%%%%%%%%%%%%%%%%%%%%%%%%%%%%%%%%%%%%%%%%%%%%%%%%%%%%%%%%%%%%%%%%%%
%%%%%%%%%%%%%%%%%%%%%%%%%%%%%%%%%%%%%%%%%%%%%%%%%%%%%%%%%%%%%%%%%%%%%%%%%%%%%

\section{Conclusions}
\label{conclusions}

In summary, we have studied the Hall effect and the magnetoresistance
close to the density-driven Mott transition within the single-band
Hubbard model in infinite dimensions. We have shown that the orbital
magnetoresistance is always nonnegative in this approximation. To
elucidate the emerging dependences on temperature, on the doping
level, and on the correlation strength, we have considered the
asymptotic regimes at very low and high temperatures analytically. We
have interpolated between these limiting cases by also having computed
all considered observables numerically. 

In the Fermi-liquid regime, we found that the correlations
suppress the magnetoresistance $\Delta\rho$ by a factor $Z^2$, where
$Z$ is the quasiparticle residue which is linear in doping close to
the Mott transition. Moreover, we have derived $\Delta\rho\sim1/T^2$. 
We found that Kohler's rule is obeyed only in this
low-temperature regime.

For temperatures greater than the half bandwidth $D$ but much smaller
than the Mott-Hubbard gap (which, therefore, was assumed to be
infinite), we have obtained the following analytical results: First,
the zero-field resistivity $\rho$ and the magnetoresistivity are
linear in temperature and diverge as $1/\delta$ close to the Mott
transition. Hence, $\Delta\rho/\rho\sim\mbox{const}$, which replaces
Kohler's rule. Second, the Hall constant always diverges as $1/\delta$
close to half filling. Finally, we have pointed out that the
high-temperature behavior of the Hall effect crucially depends on the
band structure: Generically, both the longitudinal and the Hall
conductivity go like $1/T$. This implies that the Hall constant
displays linear-in $T$ behavior and the Hall angle $\cot\theta_H$
saturates as $T\rightarrow\infty$. If, on the other hand, a
bipartite-lattice condition is satisfied, $R_H\sim\mbox{const}$ and
$\cot\theta_H\sim T$.

In the intermediate-temperature regime, we have investigated the Hall
constant, the Hall angle, and the magnetoresistance numerically. We
have argued that none of the resulting temperature dependences can
account for what is observed experimentally in the normal state of the
cuprate superconductors. On the other hand, direct comparison with
experiment has shown that the dynamical mean-field theory provides a
surprisingly accurate description of three-dimensional
transition-metal oxides \cite{Georges:1996}, and our results should
serve as a qualitative guide for what we would expect in these
systems. 

\section{Acknowledgments}

We are grateful to Gunnar P\'alsson for valuable discussions. This
work was supported by NSF, DMR  95-29138. E.L. is funded by the
Deutsche Forschungsgemeinschaft.

%%%%%%%%%%%%%%%%%%%%%%%%%%%%%%%%%%%%%%%%%%%%%%%%%%%%%%%%%%%%%%%%%%%%%%%%%%%%%
%%%%%%   
%%%%%%           FIGURE CAPTIONS
%%%%%%   
%%%%%%%%%%%%%%%%%%%%%%%%%%%%%%%%%%%%%%%%%%%%%%%%%%%%%%%%%%%%%%%%%%%%%%%%%%%%%

\newpage

%----------------------------------------------------------------------------
%----------------------------------------------------------------------------
%----------------------------------------------------------------------------
%----------------------------------------------------------------------------
%----------------------------------------------------------------------------
%----------------------------------------------------------------------------
%----------------------------------------------------------------------------
%----------------------------------------------------------------------------

%%%%%%%%%%%%%%%%%%%%%%%%%%%%%%%%%%%%%%%%%%%%%%%%%%%%%%%%%%%%%%%%%%%%%%%%%%%%
%%%%%%%%%%%%%%%%%%%%%%%%%%%%%%%%%%%%%%%%%%%%%%%%%%%%%%%%%%%%%%%%%%%%%%%%%%%%

\end{document}